%Paper: 9110009
%From: MAREK%TAUNIVM.BITNET@TAUNIVM.TAU.AC.IL
%Date: 2 Oct 91 18:56 IST

A Postscript figure is enclosed at the end of the TeX file.

=== TeX file begins here ==========================================
\catcode`\@=11 % This allows us to modify PLAIN macros.
\let\rel@x=\relax
\let\n@expand=\relax
\def\pr@tect{\let\n@expand=\noexpand}
\let\protect=\pr@tect
\let\gl@bal=\global
\newfam\cpfam
\newdimen\b@gheight             \b@gheight=12pt
\newcount\f@ntkey               \f@ntkey=0
\def\f@m{\afterassignment\samef@nt\f@ntkey=}
\def\samef@nt{\fam=\f@ntkey \the\textfont\f@ntkey\rel@x}
\def\setstr@t{\setbox\strutbox=\hbox{\vrule height 0.85\b@gheight
                                depth 0.35\b@gheight width\z@ }}

\font\fourteenrm  =cmr12 scaled\magstep1
\font\twelverm    =cmr10 scaled\magstep1
\font\tenrm       =cmr8  scaled\magstep1
\font\ninerm      =cmr7  scaled\magstep1
\font\sevenrm     =cmr6  scaled\magstep1
\font\sixrm       =cmr5  scaled\magstep1
\let\fiverm=\sixrm        % no cmr4

\font\fourteenbf  =cmbx12 scaled\magstep1
\font\twelvebf    =cmbx10 scaled\magstep1
\font\tenbf       =cmbx8  scaled\magstep1
\font\ninebf      =cmbx7  scaled\magstep1
\font\sevenbf     =cmbx6  scaled\magstep1
\font\sixbf       =cmbx5  scaled\magstep1
\let\fivebf=\sixbf        % no cmbx4
\font\seventeeni  =cmmi12 scaled\magstep2    \skewchar\seventeeni='177
\font\fourteeni   =cmmi12 scaled\magstep1     \skewchar\fourteeni='177
\font\twelvei     =cmmi10 scaled\magstep1       \skewchar\twelvei='177
\font\teni        =cmmi8  scaled\magstep1          \skewchar\teni='177
\font\ninei       =cmmi7  scaled\magstep1         \skewchar\ninei='177
\font\seveni      =cmmi6  scaled\magstep1        \skewchar\seveni='177
\font\sixi        =cmmi5  scaled\magstep1          \skewchar\sixi='177
\let\fivei=\sixi          % no cmmi4
\font\seventeensy =cmsy10 scaled\magstep3    \skewchar\seventeensy='60
\font\fourteensy  =cmsy10 scaled\magstep2     \skewchar\fourteensy='60
\font\twelvesy    =cmsy10 scaled\magstep1       \skewchar\twelvesy='60
\font\tensy       =cmsy8  scaled\magstep1          \skewchar\tensy='60
\font\ninesy      =cmsy7  scaled\magstep1         \skewchar\ninesy='60
\font\sevensy     =cmsy6  scaled\magstep1        \skewchar\sevensy='60
\font\sixsy       =cmsy5  scaled\magstep1          \skewchar\sixsy='60
\let\fivesy=\sixsy        % no cmsy4

\font\fourteenex  =cmex10 scaled\magstep2
\font\twelveex    =cmex10 scaled\magstep1
\let\tenex=\twelveex

\font\fourteensl  =cmsl12 scaled\magstep1
\font\twelvesl    =cmsl10 scaled\magstep1
\font\tensl       =cmsl8  scaled\magstep1
\let\ninesl=\ninerm     % no cmsl7

\font\fourteenit  =cmti12 scaled\magstep1
\font\twelveit    =cmti10 scaled\magstep1
\font\tenit       =cmti8  scaled\magstep1
\font\nineit      =cmti7  scaled\magstep1
\font\fourteentt  =cmtt12 scaled\magstep1
\font\twelvett    =cmtt10 scaled\magstep1
\font\tentt       =cmtt8  scaled\magstep1
\font\fourteencp  =cmcsc10 scaled\magstep2
\font\twelvecp    =cmcsc10 scaled\magstep1
\let\tencp=\twelvecp      % no cmcsc8
\def\fourteenf@nts{\relax
    \textfont0=\fourteenrm          \scriptfont0=\tenrm
      \scriptscriptfont0=\sevenrm
    \textfont1=\fourteeni           \scriptfont1=\teni
      \scriptscriptfont1=\seveni
    \textfont2=\fourteensy          \scriptfont2=\tensy
      \scriptscriptfont2=\sevensy
    \textfont3=\fourteenex          \scriptfont3=\twelveex
      \scriptscriptfont3=\tenex
    \textfont\itfam=\fourteenit     \scriptfont\itfam=\tenit
    \textfont\slfam=\fourteensl     \scriptfont\slfam=\tensl
    \textfont\bffam=\fourteenbf     \scriptfont\bffam=\tenbf
      \scriptscriptfont\bffam=\sevenbf
    \textfont\ttfam=\fourteentt
    \textfont\cpfam=\fourteencp }
\def\twelvef@nts{\relax
    \textfont0=\twelverm          \scriptfont0=\ninerm
      \scriptscriptfont0=\sixrm
    \textfont1=\twelvei           \scriptfont1=\ninei
      \scriptscriptfont1=\sixi
    \textfont2=\twelvesy          \scriptfont2=\ninesy
      \scriptscriptfont2=\sixsy
    \textfont3=\twelveex          \scriptfont3=\tenex
      \scriptscriptfont3=\tenex
    \textfont\itfam=\twelveit     \scriptfont\itfam=\nineit
    \textfont\slfam=\twelvesl     \scriptfont\slfam=\ninesl
    \textfont\bffam=\twelvebf     \scriptfont\bffam=\ninebf
      \scriptscriptfont\bffam=\sixbf
    \textfont\ttfam=\twelvett
    \textfont\cpfam=\twelvecp }
\def\tenf@nts{\relax
    \textfont0=\tenrm          \scriptfont0=\sevenrm
      \scriptscriptfont0=\fiverm
    \textfont1=\teni           \scriptfont1=\seveni
      \scriptscriptfont1=\fivei
    \textfont2=\tensy          \scriptfont2=\sevensy
      \scriptscriptfont2=\fivesy
    \textfont3=\tenex          \scriptfont3=\tenex
      \scriptscriptfont3=\tenex
    \textfont\itfam=\tenit     \scriptfont\itfam=\seveni  % no \sevenit
    \textfont\slfam=\tensl     \scriptfont\slfam=\sevenrm % no \sevensl
    \textfont\bffam=\tenbf     \scriptfont\bffam=\sevenbf
      \scriptscriptfont\bffam=\fivebf
    \textfont\ttfam=\tentt
    \textfont\cpfam=\tencp }
\def\rm{\n@expand\f@m0 }
\def\mit{\n@expand\f@m1 }         
\def\cal{\n@expand\f@m2 }
\def\it{\n@expand\f@m\itfam}
\def\sl{\n@expand\f@m\slfam}
\def\bf{\n@expand\f@m\bffam}
\def\tt{\n@expand\f@m\ttfam}
\def\caps{\n@expand\f@m\cpfam}    
\def\em@{\rel@x\ifnum\f@ntkey=0 \it \else
        \ifnum\f@ntkey=\bffam \it \else \rm \fi \fi }
\def\em{\n@expand\em@}
\def\fourteenpoint{\fourteenf@nts \samef@nt \b@gheight=14pt \setstr@t }
\def\twelvepoint{\twelvef@nts \samef@nt \b@gheight=12pt \setstr@t }
\def\tenpoint{\tenf@nts \samef@nt \b@gheight=10pt \setstr@t }
\normalbaselineskip = 20pt plus 0.2pt minus 0.1pt
\normallineskip = 1.5pt plus 0.1pt minus 0.1pt
\normallineskiplimit = 1.5pt
\newskip\normaldisplayskip
\normaldisplayskip = 20pt plus 5pt minus 10pt
\newskip\normaldispshortskip
\normaldispshortskip = 6pt plus 5pt
\newskip\normalparskip
\normalparskip = 6pt plus 2pt minus 1pt
\newskip\skipregister
\skipregister = 5pt plus 2pt minus 1.5pt
\newif\ifsingl@
\newif\ifdoubl@
\newif\iftwelv@  \twelv@true
\def\singlespace{\singl@true\doubl@false\spaces@t}
\def\doublespace{\singl@false\doubl@true\spaces@t}
\def\normalspace{\singl@false\doubl@false\spaces@t}
\def\Tenpoint{\tenpoint\twelv@false\spaces@t}
\def\Twelvepoint{\twelvepoint\twelv@true\spaces@t}
\def\spaces@t{\rel@x
      \iftwelv@ \ifsingl@\subspaces@t3:4;\else\subspaces@t1:1;\fi
       \else \ifsingl@\subspaces@t3:5;\else\subspaces@t4:5;\fi \fi
      \ifdoubl@ \multiply\baselineskip by 5
         \divide\baselineskip by 4 \fi }
\def\subspaces@t#1:#2;{
      \baselineskip = \normalbaselineskip
      \multiply\baselineskip by #1 \divide\baselineskip by #2
      \lineskip = \normallineskip
      \multiply\lineskip by #1 \divide\lineskip by #2
      \lineskiplimit = \normallineskiplimit
      \multiply\lineskiplimit by #1 \divide\lineskiplimit by #2
      \parskip = \normalparskip
      \multiply\parskip by #1 \divide\parskip by #2
      \abovedisplayskip = \normaldisplayskip
      \multiply\abovedisplayskip by #1 \divide\abovedisplayskip by #2
      \belowdisplayskip = \abovedisplayskip
      \abovedisplayshortskip = \normaldispshortskip
      \multiply\abovedisplayshortskip by #1
        \divide\abovedisplayshortskip by #2
      \belowdisplayshortskip = \abovedisplayshortskip
      \advance\belowdisplayshortskip by \belowdisplayskip
      \divide\belowdisplayshortskip by 2
      \smallskipamount = \skipregister
      \multiply\smallskipamount by #1 \divide\smallskipamount by #2
      \medskipamount = \smallskipamount \multiply\medskipamount by 2
      \bigskipamount = \smallskipamount \multiply\bigskipamount by 4 }
\def\normalbaselines{ \baselineskip=\normalbaselineskip
   \lineskip=\normallineskip \lineskiplimit=\normallineskip
   \iftwelv@\else \multiply\baselineskip by 4 \divide\baselineskip by 5
     \multiply\lineskiplimit by 4 \divide\lineskiplimit by 5
     \multiply\lineskip by 4 \divide\lineskip by 5 \fi }
\Twelvepoint  % That's the default
\interlinepenalty=50
\interfootnotelinepenalty=5000
\predisplaypenalty=9000
\postdisplaypenalty=500
\hfuzz=1pt
\vfuzz=0.2pt
\newdimen\HOFFSET  \HOFFSET=0pt
\newdimen\VOFFSET  \VOFFSET=0pt
\newdimen\HSWING   \HSWING=0pt
\dimen\footins=8in
\newskip\pagebottomfiller
\pagebottomfiller=\z@ plus \z@ minus \z@
\def\pagecontents{
   \ifvoid\topins\else\unvbox\topins\vskip\skip\topins\fi
   \dimen@ = \dp255 \unvbox255
   \vskip\pagebottomfiller
   \ifvoid\footins\else\vskip\skip\footins\footrule\unvbox\footins\fi
   \ifr@ggedbottom \kern-\dimen@ \vfil \fi }
\def\makeheadline{\vbox to 0pt{ \skip@=\topskip
      \advance\skip@ by -12pt \advance\skip@ by -2\normalbaselineskip
      \vskip\skip@ \line{\vbox to 12pt{}\the\headline} \vss
      }\nointerlineskip}
\def\makefootline{\baselineskip = 1.5\normalbaselineskip
                 \line{\the\footline}}
\newif\iffrontpage
\newif\ifp@genum
\def\nopagenumbers{\p@genumfalse}
\def\pagenumbers{\p@genumtrue}
\pagenumbers
\newtoks\paperheadline
\newtoks\paperfootline
\newtoks\letterheadline
\newtoks\letterfootline
\newtoks\letterinfo
\newtoks\date
\paperheadline={\hfil}
\paperfootline={\hss\iffrontpage\else\ifp@genum\tenrm\folio\hss\fi\fi}
\letterheadline{\iffrontpage \hfil \else
    \rm \ifp@genum page~~\folio\fi \hfil\the\date \fi}
\letterfootline={\iffrontpage\the\letterinfo\else\hfil\fi}
\letterinfo={\hfil}
\def\monthname{\rel@x\ifcase\month 0/\or January\or February\or
   March\or April\or May\or June\or July\or August\or September\or
   October\or November\or December\else\number\month/\fi}
\def\today{\monthname~\number\day, \number\year}
\date={\today}
\headline=\paperheadline % The default is
\footline=\paperfootline % \papers
\countdef\pageno=1      \countdef\pagen@=0
\countdef\pagenumber=1  \pagenumber=1
\def\advancepageno{\gl@bal\advance\pagen@ by 1
   \ifnum\pagenumber<0 \gl@bal\advance\pagenumber by -1
    \else\gl@bal\advance\pagenumber by 1 \fi
    \gl@bal\frontpagefalse  \swing@ }
\def\folio{\ifnum\pagenumber<0 \romannumeral-\pagenumber
           \else \number\pagenumber \fi }
\def\swing@{\ifodd\pagenumber \gl@bal\advance\hoffset by -\HSWING
             \else \gl@bal\advance\hoffset by \HSWING \fi }
\def\footrule{\dimen@=\prevdepth\nointerlineskip
   \vbox to 0pt{\vskip -0.25\baselineskip \hrule width 0.35\hsize \vss}
   \prevdepth=\dimen@ }
\let\footnotespecial=\rel@x
\newdimen\footindent
\footindent=24pt
\def\Textindent#1{\noindent\llap{#1\enspace}\ignorespaces}
\def\Vfootnote#1{\insert\footins\bgroup
   \interlinepenalty=\interfootnotelinepenalty \floatingpenalty=20000
   \singl@true\doubl@false\Tenpoint
   \splittopskip=\ht\strutbox \boxmaxdepth=\dp\strutbox
   \leftskip=\footindent \rightskip=\z@skip
   \parindent=0.5\footindent \parfillskip=0pt plus 1fil
   \spaceskip=\z@skip \xspaceskip=\z@skip \footnotespecial
   \Textindent{#1}\footstrut\futurelet\next\fo@t}

\def\vfootnote#1{\Vfootnote{${#1}$}}
\def\footnote#1{\attach{#1}\vfootnote{#1}}

\def\foot{\attach\footsymbolgen\vfootnote{\footsymbol}}
\let\footsymbol=\star
\newcount\lastf@@t           \lastf@@t=-1
\newcount\footsymbolcount    \footsymbolcount=0
\newif\ifPhysRev
\def\footsymbolgen{\bumpfootsymbolcount \generatefootsymbol \footsymbol }
\def\bumpfootsymbolcount{\rel@x
   \iffrontpage \bumpfootsymbolpos \else \advance\lastf@@t by 1
     \ifPhysRev \bumpfootsymbolneg \else \bumpfootsymbolpos \fi \fi
   \gl@bal\lastf@@t=\pagen@ }
\def\bumpfootsymbolpos{\ifnum\footsymbolcount <0
                            \gl@bal\footsymbolcount =0 \fi
    \ifnum\lastf@@t<\pagen@ \gl@bal\footsymbolcount=0
     \else \gl@bal\advance\footsymbolcount by 1 \fi }
\def\bumpfootsymbolneg{\ifnum\footsymbolcount >0
             \gl@bal\footsymbolcount =0 \fi
         \gl@bal\advance\footsymbolcount by -1 }
\def\fd@f#1 {\xdef\footsymbol{\mathchar"#1 }}
\def\generatefootsymbol{\ifcase\footsymbolcount \fd@f 13F \or \fd@f 279
        \or \fd@f 27A \or \fd@f 278 \or \fd@f 27B \else
        \ifnum\footsymbolcount <0 \fd@f{023 \number-\footsymbolcount }
         \else \fd@f 203 {\loop \ifnum\footsymbolcount >5
                \fd@f{203 \footsymbol } \advance\footsymbolcount by -1
                \repeat }\fi \fi }

\def\nonfrenchspacing{\sfcode`\.=3001 \sfcode`\!=3000 \sfcode`\?=3000
        \sfcode`\:=2000 \sfcode`\;=1500 \sfcode`\,=1251 }
\nonfrenchspacing
\newdimen\d@twidth
{\setbox0=\hbox{s.} \gl@bal\d@twidth=\wd0 \setbox0=\hbox{s}
        \gl@bal\advance\d@twidth by -\wd0 }
\def\removehglue{\loop \unskip \ifdim\lastskip >\z@ \repeat }
\def\roll@ver#1{\removehglue \nobreak \count255 =\spacefactor \dimen@=\z@
        \ifnum\count255 =3001 \dimen@=\d@twidth \fi
        \ifnum\count255 =1251 \dimen@=\d@twidth \fi
    \iftwelv@ \kern-\dimen@ \else \kern-0.83\dimen@ \fi
   #1\spacefactor=\count255 }
\def\step@ver#1{\rel@x \ifmmode #1\else \ifhmode
        \roll@ver{${}#1$}\else {\setbox0=\hbox{${}#1$}}\fi\fi }
\def\attach#1{\step@ver{\strut^{\mkern 2mu #1} }}
\newcount\chapternumber      \chapternumber=0
\newcount\sectionnumber      \sectionnumber=0
\newcount\equanumber         \equanumber=0
\let\chapterlabel=\rel@x
\let\sectionlabel=\rel@x
\newtoks\chapterstyle        \chapterstyle={\Number}
\newtoks\sectionstyle        \sectionstyle={\chapterlabel.\Number}
\newskip\chapterskip         \chapterskip=\bigskipamount
\newskip\sectionskip         \sectionskip=\medskipamount
\newskip\headskip            \headskip=8pt plus 3pt minus 3pt
\newdimen\chapterminspace    \chapterminspace=15pc
\newdimen\sectionminspace    \sectionminspace=10pc
\newdimen\referenceminspace  \referenceminspace=20pc
\def\chapterreset{\gl@bal\advance\chapternumber by 1
   \ifnum\equanumber<0 \else\gl@bal\equanumber=0\fi
   \sectionnumber=0 \let\sectionlabel=\rel@x
   {\pr@tect\xdef\chapterlabel{\the\chapterstyle{\the\chapternumber}}}}
\def\alphabetic#1{\count255='140 \advance\count255 by #1\char\count255}
\def\Alphabetic#1{\count255='100 \advance\count255 by #1\char\count255}
\def\Roman#1{\uppercase\expandafter{\romannumeral #1}}
\def\roman#1{\romannumeral #1}
\def\Number#1{\number #1}
\def\BLANC#1{}
\def\titleparagraphs{\interlinepenalty=9999
     \leftskip=0.03\hsize plus 0.22\hsize minus 0.03\hsize
     \rightskip=\leftskip \parfillskip=0pt
     \hyphenpenalty=9000 \exhyphenpenalty=9000
     \tolerance=9999 \pretolerance=9000
     \spaceskip=0.333em \xspaceskip=0.5em }
\def\titlestyle#1{\par\begingroup \titleparagraphs
     \iftwelv@\fourteenpoint\else\twelvepoint\fi
   \noindent #1\par\endgroup }
\def\spacecheck#1{\dimen@=\pagegoal\advance\dimen@ by -\pagetotal
   \ifdim\dimen@<#1 \ifdim\dimen@>0pt \vfil\break \fi\fi}
\def\chapter#1{\par \penalty-300 \vskip\chapterskip
   \spacecheck\chapterminspace
   \chapterreset \titlestyle{\chapterlabel.~#1}
   \nobreak\vskip\headskip \penalty 30000
   {\pr@tect\wlog{\string\chapter\space \chapterlabel}} }

\def\section#1{\par \ifnum\the\lastpenalty=30000\else
   \penalty-200\vskip\sectionskip \spacecheck\sectionminspace\fi
   \gl@bal\advance\sectionnumber by 1
   {\pr@tect
   \xdef\sectionlabel{\the\sectionstyle\the\sectionnumber}
   \wlog{\string\section\space \sectionlabel}}
   \noindent {\caps\enspace\sectionlabel.~~#1}\par
   \nobreak\vskip\headskip \penalty 30000 }
\def\subsection#1{\par
   \ifnum\the\lastpenalty=30000\else \penalty-100\smallskip \fi
   \noindent\undertext{#1}\enspace \vadjust{\penalty5000}}

\def\undertext#1{\vtop{\hbox{#1}\kern 1pt \hrule}}
\def\ACK{\par\penalty-100\medskip \spacecheck\sectionminspace
   \line{\fourteenrm\hfil ACKNOWLEDGEMENTS\hfil}\nobreak\vskip\headskip }

\def\APPENDIX#1#2{\par\penalty-300\vskip\chapterskip
   \spacecheck\chapterminspace \chapterreset \xdef\chapterlabel{#1}
   \titlestyle{APPENDIX #2} \nobreak\vskip\headskip \penalty 30000
   \wlog{\string\Appendix~\chapterlabel} }
\def\Appendix#1{\APPENDIX{#1}{#1}}
\def\appendix{\APPENDIX{A}{}}
\def\unnumberedchapters{\let\makechapterlabel=\rel@x
      \let\chapterlabel=\rel@x  \sectionstyle={\BLANC}
      \let\sectionlabel=\rel@x \sequentialequations }
\def\eqname#1{\rel@x {\pr@tect
  \ifnum\equanumber<0 \xdef#1{{\rm(\number-\equanumber)}}%
     \gl@bal\advance\equanumber by -1
  \else \gl@bal\advance\equanumber by 1
     \ifx\chapterlabel\rel@x \def\d@t{}\else \def\d@t{.}\fi
    \xdef#1{{\rm(\chapterlabel\d@t\number\equanumber)}}\fi #1}}

\def\eqn{\eqno\eqname}

\def\eqinsert#1{\noalign{\dimen@=\prevdepth \nointerlineskip
   \setbox0=\hbox to\displaywidth{\hfil #1}
   \vbox to 0pt{\kern 0.5\baselineskip\hbox{$\!\box0\!$}\vss}
   \prevdepth=\dimen@}}

\def\GENITEM#1;#2{\par \hangafter=0 \hangindent=#1
    \Textindent{$ #2 $}\ignorespaces}
\outer\def\newitem#1=#2;{\gdef#1{\GENITEM #2;}}

\newdimen\itemsize                \itemsize=30pt
\newitem\item=1\itemsize;
\newitem\sitem=1.75\itemsize;     
\newitem\ssitem=2.5\itemsize;     
\outer\def\newlist#1=#2&#3&#4;{\toks0={#2}\toks1={#3}%
   \count255=\escapechar \escapechar=-1
   \alloc@0\list\countdef\insc@unt\listcount     \listcount=0
   \edef#1{\par
      \countdef\listcount=\the\allocationnumber
      \advance\listcount by 1
      \hangafter=0 \hangindent=#4
      \Textindent{\the\toks0{\listcount}\the\toks1}}
   \expandafter\expandafter\expandafter
    \edef\c@t#1{begin}{\par
      \countdef\listcount=\the\allocationnumber \listcount=1
      \hangafter=0 \hangindent=#4
      \Textindent{\the\toks0{\listcount}\the\toks1}}
   \expandafter\expandafter\expandafter
    \edef\c@t#1{con}{\par \hangafter=0 \hangindent=#4 \noindent}
   \escapechar=\count255}
\def\c@t#1#2{\csname\string#1#2\endcsname}
\newlist\point=\Number&.&1.0\itemsize;
\newlist\subpoint=(\alphabetic&)&1.75\itemsize;
\newlist\subsubpoint=(\roman&)&2.5\itemsize;

\newcount\referencecount     \referencecount=0
\newcount\lastrefsbegincount \lastrefsbegincount=0
\newif\ifreferenceopen       \newwrite\referencewrite
\newdimen\refindent          \refindent=30pt
\def\normalrefmark#1{\attach{\scriptscriptstyle [ #1 ] }}
\let\PRrefmark=\attach
\def\NPrefmark#1{\step@ver{{\;[#1]}}}
\def\refmark#1{\rel@x\ifPhysRev\PRrefmark{#1}\else\normalrefmark{#1}\fi}
\def\refend@{\refmark{\number\referencecount}}
\def\refend{\refend@{}\space }
\def\refsend{\refmark{\count255=\referencecount
   \advance\count255 by-\lastrefsbegincount
   \ifcase\count255 \number\referencecount
   \or \number\lastrefsbegincount,\number\referencecount
   \else \number\lastrefsbegincount-\number\referencecount \fi}\space }
\def\REFNUM#1{\rel@x \gl@bal\advance\referencecount by 1
    \xdef#1{\the\referencecount }}
\def\Refnum#1{\REFNUM #1\refend@ } 
\def\REF#1{\REFNUM #1\R@FWRITE\ignorespaces}
\def\Ref#1{\Refnum #1\REFWRITE }
\def\ref{\Ref\?}
\def\REFS#1{\REFNUM #1\gl@bal\lastrefsbegincount=\referencecount
    \REFWRITE }

\def\r@fitem#1{\par \hangafter=0 \hangindent=\refindent \Textindent{#1}}
\def\refitem#1{\r@fitem{#1.}}
\def\NPrefitem#1{\r@fitem{[#1]}}
\def\NPrefs{\let\refmark=\NPrefmark \let\refitem=NPrefitem}
\def\REFWRITE{\R@FWRITE\rel@x }
\def\R@FWRITE#1{\ifreferenceopen \else \gl@bal\referenceopentrue
     \immediate\openout\referencewrite=\jobname.refs
     \toks@={\begingroup \refoutspecials \catcode`\^^M=10 }%
     \immediate\write\referencewrite{\the\toks@}\fi
    \immediate\write\referencewrite{\noexpand\refitem %
                                    {\the\referencecount}}%
    \p@rse@ndwrite \referencewrite #1}
\begingroup
 \catcode`\^^M=\active \let^^M=\relax %
 \gdef\p@rse@ndwrite#1#2{\begingroup \catcode`\^^M=12 \newlinechar=`\^^M%
         \chardef\rw@write=#1\sc@nlines#2}%
 \gdef\sc@nlines#1#2{\sc@n@line \g@rbage #2^^M\endsc@n \endgroup #1}%
 \gdef\sc@n@line#1^^M{\expandafter\toks@\expandafter{\deg@rbage #1}%
         \immediate\write\rw@write{\the\toks@}%
         \futurelet\n@xt \sc@ntest }%
\endgroup
\def\sc@ntest{\ifx\n@xt\endsc@n \let\n@xt=\rel@x
       \else \let\n@xt=\sc@n@notherline \fi \n@xt }
\def\sc@n@notherline{\sc@n@line \g@rbage }
\def\deg@rbage#1{}
\let\g@rbage=\relax    \let\endsc@n=\relax
\def\refout{\par\penalty-400\vskip\chapterskip
   \spacecheck\referenceminspace
   \ifreferenceopen \Closeout\referencewrite \referenceopenfalse \fi
   \line{\fourteenrm\hfil REFERENCES\hfil}\vskip\headskip
   \input \jobname.refs
   }
\def\refoutspecials{\sfcode`\.=1000 \interlinepenalty=1000
         \rightskip=\z@ plus 1em minus \z@ }
\def\Closeout#1{\toks0={\par\endgroup}\immediate\write#1{\the\toks0}%
   \immediate\closeout#1}
\newcount\figurecount     \figurecount=0
\newcount\tablecount      \tablecount=0
\newif\iffigureopen       \newwrite\figurewrite
\newif\iftableopen        \newwrite\tablewrite
\def\FIGNUM#1{\rel@x \gl@bal\advance\figurecount by 1
    \xdef#1{\the\figurecount}}
\def\FIGURE#1{\FIGNUM #1\F@GWRITE\ignorespaces }

\def\figitem#1{\r@fitem{#1)}}
\def\FIGWRITE{\F@GWRITE\rel@x }
\def\TABNUM#1{\rel@x \gl@bal\advance\tablecount by 1
    \xdef#1{\the\tablecount}}
\def\TABLE#1{\TABNUM #1\T@BWRITE\ignorespaces }

\def\tabitem#1{\r@fitem{#1:}}
\def\TABWRITE{\T@BWRITE\rel@x }
\def\F@GWRITE#1{\iffigureopen \else \gl@bal\figureopentrue
     \immediate\openout\figurewrite=\jobname.figs
     \toks@={\begingroup \catcode`\^^M=10 }%
     \immediate\write\figurewrite{\the\toks@}\fi
    \immediate\write\figurewrite{\noexpand\figitem %
                                 {\the\figurecount}}%
    \p@rse@ndwrite \figurewrite #1}
\def\T@BWRITE#1{\iftableopen \else \gl@bal\tableopentrue
     \immediate\openout\tablewrite=\jobname.tabs
     \toks@={\begingroup \catcode`\^^M=10 }%
     \immediate\write\tablewrite{\the\toks@}\fi
    \immediate\write\tablewrite{\noexpand\tabitem %
                                 {\the\tablecount}}%
    \p@rse@ndwrite \tablewrite #1}
\def\figout{\par\penalty-400
   \vskip\chapterskip\spacecheck\referenceminspace
   \iffigureopen \Closeout\figurewrite \figureopenfalse \fi
   \line{\fourteenrm\hfil FIGURE CAPTIONS\hfil}\vskip\headskip
   \input \jobname.figs
   }
\def\tabout{\par\penalty-400
   \vskip\chapterskip\spacecheck\referenceminspace
   \iftableopen \Closeout\tablewrite \tableopenfalse \fi
   \line{\fourteenrm\hfil TABLE CAPTIONS\hfil}\vskip\headskip
   \input \jobname.tabs
   }
\newbox\picturebox
\def\p@cht{\ht\picturebox }
\def\p@cwd{\wd\picturebox }
\def\p@cdp{\dp\picturebox }
\newdimen\xshift
\newdimen\yshift
\newdimen\captionwidth
\newskip\captionskip
\captionskip=15pt plus 5pt minus 3pt
\def\fullwidth{\captionwidth=\hsize }
\newtoks\Caption
\newif\ifcaptioned
\newif\ifselfcaptioned
\def\caption{\captionedtrue \Caption }
\newcount\linesabove
\newif\iffileexists
\newtoks\picfilename
\def\fil@#1 {\fileexiststrue \picfilename={#1}}
\def\file#1{\if=#1\let\n@xt=\fil@ \else \def\n@xt{\fil@ #1}\fi \n@xt }
\def\pl@t{\begingroup \pr@tect
    \setbox\picturebox=\hbox{}\fileexistsfalse
    \let\height=\p@cht \let\width=\p@cwd \let\depth=\p@cdp
    \xshift=\z@ \yshift=\z@ \captionwidth=\z@
    \Caption={}\captionedfalse
    \linesabove =0 \picturedefault }
\def\plot{\pl@t \selfcaptionedfalse }
\def\Picture#1{\gl@bal\advance\figurecount by 1
    \xdef#1{\the\figurecount}\pl@t \selfcaptionedtrue }

\def\s@vepicture{\iffileexists \parsefilename \redopicturebox \fi
   \ifdim\captionwidth>\z@ \else \captionwidth=\p@cwd \fi
   \xdef\lastpicture{\iffileexists
        \setbox0=\hbox{\raise\the\yshift \vbox{%
              \moveright\the\xshift\hbox{\picturedefinition}}}%
        \else \setbox0=\hbox{}\fi
         \ht0=\the\p@cht \wd0=\the\p@cwd \dp0=\the\p@cdp
         \vbox{\hsize=\the\captionwidth \line{\hss\box0 \hss }%
              \ifcaptioned \vskip\the\captionskip \noexpand\Tenpoint
                \ifselfcaptioned Figure~\the\figurecount.\enspace \fi
                \the\Caption \fi }}%
    \endgroup }
\let\endpicture=\s@vepicture
\def\savepicture#1{\s@vepicture \global\let#1=\lastpicture }
\def\displaypicture{\fullwidth \s@vepicture $$\lastpicture $${}}
\def\toppicture{\fullwidth \s@vepicture \topinsert
    \lastpicture \medskip \endinsert }
\def\midpicture{\fullwidth \s@vepicture \midinsert
    \lastpicture \endinsert }
\def\leftpicture{\pres@tpicture
    \dimen@i=\hsize \advance\dimen@i by -\dimen@ii
    \setbox\picturebox=\hbox to \hsize {\box0 \hss }%
    \wr@paround }
\def\rightpicture{\pres@tpicture
    \dimen@i=\z@
    \setbox\picturebox=\hbox to \hsize {\hss \box0 }%
    \wr@paround }
\def\pres@tpicture{\gl@bal\linesabove=\linesabove
    \s@vepicture \setbox\picturebox=\vbox{
         \kern \linesabove\baselineskip \kern 0.3\baselineskip
         \lastpicture \kern 0.3\baselineskip }%
    \dimen@=\p@cht \dimen@i=\dimen@
    \advance\dimen@i by \pagetotal
    \par \ifdim\dimen@i>\pagegoal \vfil\break \fi
    \dimen@ii=\hsize
    \advance\dimen@ii by -\parindent \advance\dimen@ii by -\p@cwd
    \setbox0=\vbox to\z@{\kern-\baselineskip \unvbox\picturebox \vss }}
\def\wr@paround{\Caption={}\count255=1
    \loop \ifnum \linesabove >0
         \advance\linesabove by -1 \advance\count255 by 1
         \advance\dimen@ by -\baselineskip
         \expandafter\Caption \expandafter{\the\Caption \z@ \hsize }%
      \repeat
    \loop \ifdim \dimen@ >\z@
         \advance\count255 by 1 \advance\dimen@ by -\baselineskip
         \expandafter\Caption \expandafter{%
             \the\Caption \dimen@i \dimen@ii }%
      \repeat
    \edef\n@xt{\parshape=\the\count255 \the\Caption \z@ \hsize }%
    \par\noindent \n@xt \strut \vadjust{\box\picturebox }}
\let\picturedefault=\relax
\let\parsefilename=\relax
\def\redopicturebox{\let\picturedefinition=\rel@x
   \errhelp=\disabledpictures
   \errmessage{This version of TeX cannot handle pictures.  Sorry.}}
\newhelp\disabledpictures
     {You will get a blank box in place of your picture.}
\def\FRONTPAGE{\ifvoid255\else\vfill\penalty-20000\fi
   \gl@bal\pagenumber=1     \gl@bal\chapternumber=0
   \gl@bal\equanumber=0     \gl@bal\sectionnumber=0
   \gl@bal\referencecount=0 \gl@bal\figurecount=0
   \gl@bal\tablecount=0     \gl@bal\frontpagetrue
   \gl@bal\lastf@@t=0       \gl@bal\footsymbolcount=0}

\def\papers{\papersize\headline=\paperheadline\footline=\paperfootline}
\def\papersize{\hsize=35pc \vsize=50pc \hoffset=0pc \voffset=1pc
   \advance\hoffset by\HOFFSET \advance\voffset by\VOFFSET
   \pagebottomfiller=0pc
   \skip\footins=\bigskipamount \normalspace }
\papers  %  This is the default
\newskip\lettertopskip       \lettertopskip=20pt plus 50pt
\newskip\letterbottomskip    \letterbottomskip=\z@ plus 100pt
\newskip\signatureskip       \signatureskip=40pt plus 3pt
\def\lettersize{\hsize=6.5in \vsize=8.5in \hoffset=0in \voffset=0.5in
   \advance\hoffset by\HOFFSET \advance\voffset by\VOFFSET
   \pagebottomfiller=\letterbottomskip
   \skip\footins=\smallskipamount \multiply\skip\footins by 3
   \singlespace }
\def\MEMO{\lettersize \headline=\letterheadline \footline={\hfil }%
   \let\rule=\memorule \FRONTPAGE \memohead }

\def\memodate{\afterassignment\MEMO \date }
\def\memit@m#1{\smallskip \hangafter=0 \hangindent=1in
    \Textindent{\caps #1}}
\def\subject{\memit@m{Subject:}}
\def\topic{\memit@m{Topic:}}
\def\from{\memit@m{From:}}
\def\to{\rel@x \ifmmode \rightarrow \else \memit@m{To:}\fi }
\def\memorule{\medskip\hrule height 1pt\bigskip}  % default definitions
\def\memohead{\centerline{\fourteenrm MEMORANDUM}}% see phyzzx.local
\newwrite\labelswrite
\newtoks\rw@toks
\def\letters{\lettersize
   \headline=\letterheadline \footline=\letterfootline
   \immediate\openout\labelswrite=\jobname.lab}

\let\letterhead=\rel@x
\def\addressee#1{\medskip\line{\hskip 0.75\hsize plus\z@ minus 0.25\hsize
                               \the\date \hfil }%
   \vskip \lettertopskip
   \ialign to\hsize{\strut ##\hfil\tabskip 0pt plus \hsize \crcr #1\crcr}
   \writelabel{#1}\medskip \noindent\hskip -\spaceskip \ignorespaces }
\def\rwl@begin#1\cr{\rw@toks={#1\crcr}\rel@x
   \immediate\write\labelswrite{\the\rw@toks}\futurelet\n@xt\rwl@next}
\def\rwl@next{\ifx\n@xt\rwl@end \let\n@xt=\rel@x
      \else \let\n@xt=\rwl@begin \fi \n@xt}
\let\rwl@end=\rel@x
\def\writelabel#1{\immediate\write\labelswrite{\noexpand\labelbegin}
     \rwl@begin #1\cr\rwl@end
     \immediate\write\labelswrite{\noexpand\labelend}}
\newtoks\FromAddress         \FromAddress={}
\newtoks\sendername          \sendername={}
\newbox\FromLabelBox
\newdimen\labelwidth          \labelwidth=6in
\def\makelabels{\afterassignment\Makelabels \sendersname=}
\def\Makelabels{\FRONTPAGE \letterinfo={\hfil } \MakeFromBox
     \immediate\closeout\labelswrite  \input \jobname.lab\vfil\eject}
\let\labelend=\rel@x
\def\labelbegin#1\labelend{\setbox0=\vbox{\ialign{##\hfil\cr #1\crcr}}
     \MakeALabel }
\def\MakeFromBox{\gl@bal\setbox\FromLabelBox=\vbox{\Tenpoint
     \ialign{##\hfil\cr \the\sendername \the\FromAddress \crcr }}}
\def\MakeALabel{\vskip 1pt \hbox{\vrule \vbox{
        \hsize=\labelwidth \hrule\bigskip
        \leftline{\hskip 1\parindent \copy\FromLabelBox}\bigskip
        \centerline{\hfil \box0 } \bigskip \hrule
        }\vrule } \vskip 1pt plus 1fil }
\def\signed#1{\par \nobreak \bigskip \dt@pfalse \begingroup
  \everycr={\noalign{\nobreak
            \ifdt@p\vskip\signatureskip\gl@bal\dt@pfalse\fi }}%
  \tabskip=0.5\hsize plus \z@ minus 0.5\hsize
  \halign to\hsize {\strut ##\hfil\tabskip=\z@ plus 1fil minus \z@\crcr
          \noalign{\gl@bal\dt@ptrue}#1\crcr }%
  \endgroup \bigskip }

\newbox\letterb@x
\def\lettertext{\par \vskip\parskip \unvcopy\letterb@x \par }
\def\multiletter{\setbox\letterb@x=\vbox\bgroup
      \everypar{\vrule height 1\baselineskip depth 0pt width 0pt }
      \singlespace \topskip=\baselineskip }
\def\letterend{\par\egroup}
\newskip\frontpageskip
\newtoks\Pubnum   
\newtoks\Pubtype  \let\pubtype=\Pubtype
\newif\ifp@bblock  \p@bblocktrue
\def\PH@SR@V{\doubl@true \baselineskip=24.1pt plus 0.2pt minus 0.1pt
             \parskip= 3pt plus 2pt minus 1pt }
\def\PHYSREV{\papers\PhysRevtrue\PH@SR@V}

\def\titlepage{\FRONTPAGE\papers\ifPhysRev\PH@SR@V\fi
   \ifp@bblock\p@bblock \else\hrule height\z@ \rel@x \fi }
\def\nopubblock{\p@bblockfalse}
\def\endpage{\vfil\break}
\frontpageskip=12pt plus .5fil minus 2pt
\Pubtype={}
\Pubnum={}
\def\p@bblock{\begingroup \tabskip=\hsize minus \hsize
   \baselineskip=1.5\ht\strutbox \topspace-2\baselineskip
   \halign to\hsize{\strut ##\hfil\tabskip=0pt\crcr
       \the\Pubnum\crcr\the\date\crcr\the\pubtype\crcr}\endgroup}
\def\title#1{\vskip\frontpageskip \titlestyle{#1} \vskip\headskip }
\def\author#1{\vskip\frontpageskip\titlestyle{\twelvecp #1}\nobreak}

\def\address#1{\par\kern 5pt\titlestyle{\twelvepoint\it #1}}
\def\andaddress{\par\kern 5pt \centerline{\sl and} \address}

\def\abstract{\par\dimen@=\prevdepth \hrule height\z@ \prevdepth=\dimen@
   \vskip\frontpageskip\centerline{\fourteenrm ABSTRACT}\vskip\headskip }

\def\\{\rel@x \ifmmode \backslash \else {\tt\char`\\}\fi }
\def\sequentialequations{\rel@x \if\equanumber<0 \else
  \gl@bal\equanumber=-\equanumber \gl@bal\advance\equanumber by -1 \fi }
\def\nextline{\unskip\nobreak\hfill\break}

\def\journal#1&#2(#3){\begingroup \let\journal=\dummyj@urnal
    \unskip, \sl #1\unskip~\bf\ignorespaces #2\rm
    (\afterassignment\j@ur \count255=#3), \endgroup\ignorespaces }
\def\j@ur{\ifnum\count255<100 \advance\count255 by 1900 \fi
          \number\count255 }
\def\dummyj@urnal{%
    \toks@={Reference foul up: nested \journal macros}%
    \errhelp={Your forgot & or ( ) after the last \journal}%
    \errmessage{\the\toks@ }}
\def\cropen#1{\crcr\noalign{\vskip #1}}
\def\crr{\cropen{3\jot }}
\def\topspace{\hrule height 0pt depth 0pt \vskip}

\def\Buildrel#1\under#2{\mathrel{\mathop{#2}\limits_{#1}}}
\def\becomes#1{\mathchoice{\becomes@\scriptstyle{#1}}
   {\becomes@\scriptstyle{#1}} {\becomes@\scriptscriptstyle{#1}}
   {\becomes@\scriptscriptstyle{#1}}}
\def\becomes@#1#2{\mathrel{\setbox0=\hbox{$\m@th #1{\,#2\,}$}%
        \mathop{\hbox to \wd0 {\rightarrowfill}}\limits_{#2}}}

\def\Tr{\mathop{\rm Tr}\nolimits}

\let\int=\intop         
\def\lsim{\mathrel{\mathpalette\@versim<}}
\def\gsim{\mathrel{\mathpalette\@versim>}}
\def\@versim#1#2{\vcenter{\offinterlineskip
        \ialign{$\m@th#1\hfil##\hfil$\crcr#2\crcr\sim\crcr } }}
\def\big#1{{\hbox{$\left#1\vbox to 0.85\b@gheight{}\right.\n@space$}}}
\def\Big#1{{\hbox{$\left#1\vbox to 1.15\b@gheight{}\right.\n@space$}}}
\def\bigg#1{{\hbox{$\left#1\vbox to 1.45\b@gheight{}\right.\n@space$}}}
\def\Bigg#1{{\hbox{$\left#1\vbox to 1.75\b@gheight{}\right.\n@space$}}}
\def\){\mskip 2mu\nobreak }
\let\sec@nt=\sec
\def\sec{\rel@x\ifmmode\let\n@xt=\sec@nt\else\let\n@xt\section\fi\n@xt}
\def\obsolete#1{\message{Macro \string #1 is obsolete.}}
\def\firstsec#1{\obsolete\firstsec \section{#1}}
\def\firstsubsec#1{\obsolete\firstsubsec \subsection{#1}}
\def\thispage#1{\obsolete\thispage \gl@bal\pagenumber=#1\frontpagefalse}
\def\thischapter#1{\obsolete\thischapter \gl@bal\chapternumber=#1}
\def\splitout{\obsolete\splitout\rel@x}
\def\prop{\obsolete\prop \propto }
\def\nextequation#1{\obsolete\nextequation \gl@bal\equanumber=#1
   \ifnum\the\equanumber>0 \gl@bal\advance\equanumber by 1 \fi}
\def\BOXITEM{\afterassigment\B@XITEM\setbox0=}
\def\B@XITEM{\par\hangindent\wd0 \noindent\box0 }
\def\phyzzx{PHY\setbox0=\hbox{Z}\copy0 \kern-0.5\wd0 \box0 X}
        
\everyjob{\xdef\today{\monthname~\number\day, \number\year}
                          }
\message{ by V.K.}
\catcode`\@=12 % at signs are no longer letters
% \hsize=17truecm
% \vsize=23truecm
% \hoffset=-0.8truecm
% \voffset=-2.0truecm
\overfullrule=0pt
\sequentialequations

\catcode`\@=11 % This allows us to modify PLAIN macros.

\def\myeqname#1#2{\rel@x {\pr@tect
  \ifnum\equanumber<0 \xdef#1{{\rm(\number-\equanumber#2)}}%
     \gl@bal\advance\equanumber by -1
  \else \gl@bal\advance\equanumber by 1
     \ifx\chapterlabel\rel@x \def\d@t{}\else \def\d@t{.}\fi
    \xdef#1{{\rm(\chapterlabel\d@t\number\equanumber#2)}}\fi #1}}

\def\eqreset{\rel@x {\pr@tect
  \ifnum\equanumber<0    \gl@bal\advance\equanumber by 1
  \else \gl@bal\advance\equanumber by -1\fi}}

\catcode`\@=12 % at signs are no longer letters

\def\myeqn#1#2{\eqno{\myeqname{#1}{#2}}}

\def\ZZ{\hbox{$
\not
\kern0.15em\not
\kern-0.21em\lower0.2em
\vbox{\hrule width 0.52em height 0.06em depth 0pt}
\kern-0.50em\raise0.7em
\vbox{\hrule width 0.52em height 0.06em depth 0pt}
$}}

\def\({[}
\def\){]}
\def\comm#1{\nextline{\tt #1}\nextline}
\def\comm#1{\relax}
\def\alphaprime{\hbox{$(\alpha^{\prime})^{-1}$}}

\def\ee{{=}}

\def\kN{\hbox{$(k=1,2,\dots,N_c)$}}
\def\MM{\hbox{$M$}}
\def\MMM{\hbox{${\cal M}$}}
\def\nn{n} % instead of n_1, can change back to n_1
\def\O{{\cal O}}
\def\sumk{\sum_{k=1}^{N_c}}
\def\suml{\sum_{l=1}^{N_c}}
\def\Tr{\,\hbox{Tr}\,}

\REF\NQM{For a review see:
J. J. J. Kokkedee, {\sl The Quark Model},
W. A. Benjamin, New York, 1969.}

\REF\CA{For a detailed exposition of the current algebra see:
\nextline
S. L. Adler and R.F. Dashen,
{\sl  Current Algebras and Applications to Particle Physics},
W.~A.~Benjamin, New York, 1968;\nextline
S.B. Treiman, R. Jackiw and D.J. Gross,
{\sl Lectures on Current Algebra and its Applications},
Princeton Univ. Press, 1972;\nextline
S.B. Treiman, R. Jackiw, B. Zumino and E. Witten,
{\sl Current Algebra and Anomalies},
Princeton Univ. Press, 1985.}

 \REF\GOR{M. Gell-Mann, R. J. Oakes and B. Renner \journal Phys. Rev.
 &175(68)2195, and references therein.}

\REF\GL{J. Gasser and H. Leutwyler
\journal Phys. Rep. &87(82)77.}

\REF\GM{H. Georgi and A. Manohar \journal Nucl. Phys. &B310(88)527.}

\REF\KaplanI{
D. B. Kaplan \journal Phys. Lett. &B235(90)163.}

\REF\Weinberg{S. Weinberg  \journal Phys. Rev. Lett. &65(90)1181.}

\REF\KaplanII{D. B. Kaplan \journal Nucl. Phys. &B351(91)137.}

\REF\StechNRQM{U. Ellwanger and  B. Stech
\journal Z. Phys. &C49(91)683.}

\REF\tH{G. `t Hooft \journal Nucl. Phys. &B72(74)461,
{\sl ibid} \journal&B75(74)461.}

\REF\CCG{C. G. Callan, N. Coote and D. J. Gross
\journal Phys. Rev. &D13(76)1649.}

\REF\Einhorn{M. Einhorn
\journal Phys. Rev. &D14(76)3451.}

\REF\BESW{R.C. Brower, J. Ellis, M.G. Schmidt and J.H. Weis
\journal Nucl. Phys. &B128(77)131, {\sl ibid}
\journal {} &B128(77)175.}

\REF{\WittenBoson}{E. Witten \journal Comm. Math. Phys. &92(1984)455.}

\REF{\Gonzales}{D. Gonzales and A.N. Redlich
\journal Nucl. Phys. &B256(1985)621.}

\REF{\DFS}{G.D. Date, Y. Frishman and J. Sonnenschein,
{\sl Nucl. Phys.} {\bf B283}(1987),365.}

\REF{\FS}{Y. Frishman and J. Sonnenschein,
{\sl Nucl. Phys.} {\bf B294}(1987),801.}

\REF{\FZ}{Multi-baryons are discussed in
Y.~Frishman and W.~J.~Zakrzewski
\journal Nucl. Phys. &B331(1990)781 and in {\sl Proc.
of the Int. High Energy Physics Conference}, Singapore(1990).}

\REF\FK{Y. Frishman and M. Karliner
\journal Nucl. Phys. &B334(1990)393.}

\REF\HBP{The one flavor case in the discretized light-cone
formalism is discussed in
K.~Hornbostel, S.~J.~Brodsky and H.C.~Pauli
\journal Phys. Rev. &D41(1990)3814.}

\REF{\Skyrme}{T.H.R. Skyrme \journal Proc. Roy. Soc. London
&A260(1961)127.}

\REF{\SM}{E. Witten \journal   Nucl. Phys. &B223(1983)422,
{\sl ibid}  433;
G. Adkins, C. Nappi and \hbox{E. Witten}
 \journal Nucl. Phys. &B228(1983)433;
for the 3 flavor extension of the model see:
E. Guadagnini \journal Nucl. Phys. &236(1984)35;
P. O. Mazur, M. A. Nowak and \hbox{M. Prasza\l owicz},
 \sl Phys. Lett. \rm {\bf147B}(1984),137.}

\REF\WittenLewes{E. Witten in Lewes Workshop Proc.; A. Chodos
{\sl et al.}, Eds; Singapore, World Scientific, 1984.}

\REF\QED{J. Frohlich, E. Seiler
\journal Helv. Phys. Acta &49(76)889.}

\REF\DJ{L. Dolan and R. Jackiw
\journal Phys. Rev. &D9(74)3320.}

\REF\BLS{W. A. Bardeen, B. W. Lee and  R. E. Shrock
\journal Phys. Rev. &D14(76)985.}

\REF\BG{P. Binetruy, M. K. Gaillard
\journal Phys. Rev. &D32(85)931.}

\REF\GLtemp{J. Gasser and H. Leutwyler
\journal Phys. Lett. &184B(87)83;
\sl Phys. Lett. \rm {\bf 188B}(1987),477.}

\REF\CEOI{B.A. Campbell, J. Ellis and K.A. Olive
\journal Phys. Lett. &B235(90)325.}
\REF\CEOII{B.A. Campbell, J. Ellis and K.A. Olive
\journal Nucl. Phys. &B345(90)57.}

\REF\EFKb{J. Ellis, Y. Frishman and M. Karliner,
work in progress.}

\REF\EFHK{J. Ellis, Y. Frishman A. Hannani and M. Karliner,
work in progress.}

\frontpagetrue
{\baselineskip 15pt
\null
\line{\hfill CERN-TH-6218/91}
\line{\hfill WIS-91/61-PH AUG}
\line{\hfill TAUP-1906-91}

\vskip 0.9cm
\centerline{\bf \fourteenpoint Constituent quarks as solitons}
\vskip 0.9cm
\centerline{\bf John Ellis\foot{\rm e-mail: JOHNE@CERNVM.BITNET}}
\centerline{CERN -- Geneva}
\vskip .4cm
\centerline{\bf Yitzhak Frishman\foot{\rm e-mail: FNFRISHM@WEIZMANN.BITNET}}
\centerline {Department of Physics, Weizmann Institute of Science }
\centerline {76100 Rehovot, Israel}
\vskip .4cm
\centerline{and}
\vskip .4cm
\centerline{\bf Marek Karliner\foot{\rm e-mail: MAREK@TAUNIVM.BITNET}}
\centerline{Raymond and Beverly Sackler Faculty of Exact Sciences}
\centerline{School of Physics and Astronomy}
\centerline{Tel-Aviv University, 69978 Tel-Aviv, Israel.}
\vskip 0.5cm

\abstract
We exhibit soliton solutions of QCD in two dimensions
that have the quantum numbers of quarks. They exist only for quarks
heavier than the dimensional gauge coupling, and have infinite
energy, corresponding to the presence of a string carrying
the non-singlet color flux off to spatial infinity.
The quark solitons also disappear at finite temperature,
as the temperature-dependent effective quark mass is reduced
in the approach to the quark/hadron phase transition.

\vskip 1.2cm
\vfill
\line{CERN-TH-6218/91 \hfill}
\line{September 1991 \hfill}
} % end of small baselineskip
\endpage
\pagenumber=1

\chapter{Introduction}
One of the key problems in non-perturbative QCD is the understanding
of ``constituent" quarks from first principles. Historically, the
first indications for the physical reality of quarks came from the
successes of the constituent quark model in light hadron
spectroscopy and matrix elements, assuming $m_{u,d}\simeq 300$
MeV and $m_s\simeq500$ MeV.\refmark{\NQM}
Subsequently, current algebra was abstracted from the lagrangian
for the light quarks $(u,d,s)$ and applied successfully to
calculate the properties and interactions of pions and
kaons.\refmark{\CA}
However, the success of quark current algebra
could be understood only in the context of approximate chiral
symmetry, according to which the $u$, $d$ and $s$ masses in the
lagrangian were much smaller that the original ``constituent"
masses.\refmark{\GOR}
The so-called ``current quark" masses are a few MeV for
$u$ and $d$ and ${\cal O}(100$ to $200)$ MeV for the
$s$ quark.\refmark{\GL}
Thus the question arises:
what is the relationship between ``current" and ``constituent"
quarks in QCD? In the cases of the ``heavy" quarks
$(c$, $b$, $t)$, the difference between ``constituent"
and ``current" quarks is presumably not so great, although even
the successful quarkonium potential model has not been derived
completely from QCD. Even more of a
challenge\refmark{\GM-\StechNRQM}
is to derive the
constituent versions of the light
$u$, $d$ and $s$ quarks from QCD.

Refs.~[\KaplanI],[\KaplanII] proposed that the constituent quarks
are soliton  solutions of a hypothetical  effective lagrangian of
QCD, with  effective bosonic  fields carrying both  color and
flavor.  It  is not  known whether  such an  effective lagrangian
exists in a  well defined sense, nor is it  known whether there are
stable, non-trivial  classical   solutions.    Sidestepping  these
questions,   ref.~[\KaplanII]  makes the interesting
observation   that  the   conjectured
group-theoretical structure favors  solitons with quantum numbers
of  quarks.   In  the  absence  of  a  well-justified  effective
dynamics of QCD, however, further progress requires a
theoretical laboratory where the relevant phenomena are explicitly
calculable.

A useful laboratory for studying this problem is QCD in 2
dimensions.\refmark{\tH-\HBP}
This theory can be written in bosonized form\refmark{\WittenBoson}
for
arbitrary numbers of colors $N_c$ and flavors
$N_f$.\refmark{\DFS}
It reflects accurately the phenomena of quark confinement
and condensation in the vacuum that we expect to occur in QCD
in 4 dimensions. Even though Goldstone bosons cannot exist in
2 dimensions, QCD$_2$ yields a relation between the pion mass,
light quark mass and quark condensate similar to that expected on
the basis of approximate chiral symmetry in QCD$_4$.\refmark{\CCG}
Moreover, QCD$_2$ has finite-energy soliton solutions for
arbitrary values of $N_c$ and $N_f$ that can be interpreted as
baryons,\refmark{\DFS-\FK}
 in close analogy with the Skyrmion interpretation
of baryons as solitons in QCD$_4$.\refmark{\Skyrme{-}\WittenLewes}

In this paper we start an investigation of constituent quarks
in QCD$_2$. Specifically, we examine QCD$_2$ with one heavy
quark flavor, and look for soliton solutions
corresponding to a single quark, i.e. baryon number
1 (the baryon number is normalized to be $N_c$ for the nucleon).
They exist, for a sufficiently heavy quark $Q$,
but have infinite energy, corresponding to
a string carrying the non-singlet color flux off to spatial
infinity. These quark soliton solutions disappear when the meson
mass parameter \MM\
is reduced to become comparable to the gauge
coupling strength $e_c$ (which, we recall, has the dimension of
mass in QCD$_2$).They disappear when it is energetically favoured
to create a new $\bar{Q} Q$ pair, rather than to create a string
with tension $\alphaprime\sim e_c^2$
(for the analogous result in QED$_2$, see ref.~[\QED]).
The quark solitons also disappear when the temperature $T$ is
increased to ${\cal O}(\MM)$, because of a reduction in the
$T$-dependent effective meson mass. The quark condensate and
the baryonic solitons disappear at a somewhat higher temperature,
signalling deconfinement at the quark/hadron phase transition.

\chapter{Quarks as Solitons}
We start with the bosonized\refmark{\WittenBoson}
effective action for QCD$_2$\refmark{\DFS,\FS}
$$
\left\{
\eqalign{
& A(N_F{=}1) = S[h] + {1\over 2}\int d^2x(\partial_\mu\phi)^2\cr
& + \MM^2 \int d^2 x
\Tr \left[ h  e^{ i\sqrt{{4\pi\over N_c}}\phi}
+ h^{\dagger} e^{-i\sqrt{{4\pi\over N_c}}\phi}\right] \cr
& - {e^2_c\over 2\pi} \int d^2x \Tr H^2\cr}
\right.\eqn\SI$$
where we took one flavor $N_f{=}1$ for simplicity,
$N_c$ is the number of colors, $h$
is the SU($N_c$) color matrix of field operators, and
$$S[h] = {1\over8\pi} \int d^2 x
\Tr\left( \partial_\mu h \partial^\mu h^{\dagger}\right)
+ {1\over 12 \pi} {\int}_B d^3 y \epsilon^{ijk}
\Tr
\left(h^{\dagger} \partial_i h\right)
\left(h^{\dagger} \partial_j h\right)
\left(h^{\dagger} \partial_k h\right)
\eqn\SIa$$
where the last term is the Wess-Zumino term.
In eq.~\SI\
 $e_c$ is the dimensional gauge coupling and $\phi$
is the baryon number field.  The mass scale \MM\ is related to the
current quark mass $m_Q$ and to the gauge coupling,
in the limit where $e_c \gg m_Q$,
 by\refmark{\DFS}
$$\MM=[C N_c\, m_Q({e_c\over\sqrt\pi})^p]^{{1\over p+1}} ;\qquad
p=1-{1\over N_c} \myeqn{\SIIa}{a}$$
where $C={1\over2} e^{\gamma} \simeq 0.89$,
and $\gamma $ is Euler's constant.
We will be interested in
the opposite limit $m_Q \gg e_c$, hence we choose
$\MM{=}m_Q$.
The field $H$ in \SI\ is related to $h$ through
\eqreset
$$\partial_-H = i h\partial_-h^+ \myeqn{\SIIb}{b}$$
We define hermitian fields $\varphi$ by
$$h = e^{i\sqrt {4\pi}\varphi}\eqn\SIII$$
and take
$\Tr\varphi=0$,
absorbing a possible trace term
$\Tr\varphi = {\displaystyle 2\pi n\over\displaystyle\sqrt {4\pi}}$
in $\phi$.

To search for the lowest-energy classical solutions,\refmark{\DFS}
we take $\varphi$ diagonal
\comm{??? diagonal $\rarrow$ lowest energy}
with entries $\varphi_k$, \kN.
$$\left\{
\eqalign{
& A=\int\bigg[{1\over 2}\sumk(\partial_\mu\varphi_k)^2 + {1\over
2}(\partial_\mu\phi)^2\bigg]\cr
& +2\MM^2\sumk\int \cos\bigg (\sqrt {4\pi}\varphi_k
+\sqrt {{4\pi\over N_c}}\phi\bigg ) -
2e^2_c\sumk\int\varphi_k^2\cr}\right.\eqn\SIV$$
Defining the shifted variables
$$\chi_k = \varphi_k + {1\over\sqrt {N_c}}\phi\qquad\kN
\eqn\SV$$
we find that for static $\chi_k$ the Hamiltonian takes the form
$$\left\{\eqalign{&H(\chi_k)
= {1\over 2}\sumk\int (\partial_1\chi_k)^2 + V \cr
&V = 2\MM^2\sumk\bigg[1- \cos\sqrt {4\pi}\chi_k\bigg]
\,+\, 2e^2_c\sumk\chi_k^2 \,-\,
{2e^2_c\over N_c}\bigg (\suml\chi_l\bigg )^2\cr}\right.
\eqn\SVI$$
and the baryon number is\refmark{\DFS,\FS}
$$B = {1\over\sqrt\pi}\sumk[\chi_k(\infty) - \chi_k(-\infty)]
\eqn\SVII$$
The field equations for the static case are
$$\chi_k'' - 4\MM^2\sqrt\pi \,\sin\sqrt {4\pi}
\chi_k - 4e^2_c\bigg[\chi_k -
{1\over N_c}\suml\chi_l\bigg] = 0 \qquad\kN
\eqn\SVIII$$
Let us look for solutions where $\chi_k(-\infty)=0$ for all $k$.
Then, at $x=\infty$, we must have
$$\MM^2\sqrt\pi\, \sin\sqrt {4\pi}\chi_k(\infty)
+ e_c^2\bigg[\chi_k(\infty) -
{1\over N_c}\suml\chi_l(\infty)\bigg] = 0
\eqn\SIX$$
If we seek a solution with
$\chi_k(\infty)\equiv\chi(\infty)$ for all $k$,
we find that
$\chi_k(\infty)={1\over 2}\sqrt\pi n$,
and
\hbox{$B={1\over 2}n N_c$.}
But to have the eigenvalues of the squared mass-matrix
$\left[ \partial^2V/\partial\chi_k\partial\chi_l\right] $
all
positive, we must have even $n$, and thus integer baryon number
$B\ee N_c$, corresponding to a nucleon state.

In our search for quark solitons, let us first consider
the simplest case $N_c=2$.
Then, the equations \SIX\ become
$$\sin \sqrt{4\pi} \chi_1(\infty) = - \epsilon \sqrt{\pi}
\left( \chi_1(\infty)-\chi_2(\infty)\right)
\myeqn{\SIXa}{a}$$
\eqreset
$$\sin \sqrt{4\pi} \chi_2(\infty) = - \epsilon \sqrt{\pi}
\left( \chi_2(\infty)-\chi_1(\infty)\right)
\myeqn{\SIXb}{b}$$
where
$\epsilon\equiv e_c^2/2\pi \MM^2$.
Looking at the equations \SIXa,\SIXb, we see that we can have
$$\chi_2(\infty)=-\chi_1(\infty)+\nn\sqrt\pi\myeqn{\SXa}{a}$$
or
\eqreset
$$\chi_2(\infty)=\chi_1(\infty)+(\nn+{1\over 2})\sqrt\pi\myeqn{\SXb}{b}$$
In case (a) we have $B{=}\nn$, and in case (b) we have
$B={\displaystyle 2\over\displaystyle
\sqrt\pi}\,\chi_1(\infty)+(\nn+{1\over 2})$.
However, looking again at the
eigenvalues of the matrix of second derivatives of the potential,
and requiring
positivity, we exclude case (b) (which otherwise would have
 non-integer baryon number).
\par
For case (a), where $B=\nn$, we may have quarks for $\nn=1$,
and other non-baryonic
solitons for odd values of $\nn$.
\comm{??? what does "non-baryonic" refer to?}
As seen in the figure,
there is no such solution when $\epsilon=e^2_c/2\pi \MM^2>1$
(for the analogous result in QED$_2$, see ref.~[\QED]).
This can also be seen directly from the fact that
$\left\vert \sin(x)/x \right\vert \le 1$ for any $x$.
For $\epsilon \ll 1$, however,
 we find a series
of solutions with positive second derivative matrix.\foot{See
ref.~[\BLS] for comparison.}
Defining
$$\xi = \sqrt{ 4\pi}\bigg[\chi_1(\infty)
- {1\over 2}\nn\sqrt\pi\bigg]\eqn\SXI$$
these are
$$\xi_l =\bigg[\pi - {e^2_c\over2 \MM^2}\bigg]
\left\{\matrix{&(2l)~{\rm
for}~\nn~{\rm
even}\cr &(2l+1)~{\rm for}~\nn~{\rm odd}\cr}\right\}\eqn\SXII$$
in the limit where
\ \hbox{$l{\displaystyle e^2_c\over\displaystyle \MM^2}\ll{\displaystyle
\pi\over \displaystyle 2}$.}
The number of solutions increases as
$\epsilon = e_c^2/2\pi \MM^2$ decreases, as seen in the figure, and
one can find solutions numerically for
the general case $\epsilon < 1$.
The solutions \SXII\ above have infinite energy, with classical
string tension
$$\alphaprime
\approx \pi e^2_c\left\{\matrix{&(2l)^2~{\rm for}~\nn~{\rm even}\cr
&(2l+1)^2~{\rm for}~\nn~{\rm odd}\cr}\right.\eqn\SXIII$$
They correspond to excitations of ``colored" states.  The case
\hbox{$\nn{=}1,~(2l{+}1){=} 1$} is the single constituent quark
soliton.
\par
Note that the string tension \SXIII\
grows like $l^2$ for even $\nn$, and like
($l+{1\over 2})^2$ for odd $\nn$.
A possible interpretation is that the classical string
tension is growing like $I^2$, where $I$
is the color isospin of the state:
quantum mechanically it obviously has to become $I(I+1)$.

\chapter{Generalization to $N_c >2$}

It is a simple matter to generalize the above asymptotic
solution \SXa\ to an arbitrary number of colors. The
equations \SIX\ clearly require\break
\hbox{$\sqrt{4\pi}\chi_k(\infty)= n_k\pi + {\cal O}
\left( e_c^2/\sqrt{\pi}\MM^2\right) $}
for some integers
$\{n_k\}$. The positivity of the eigenvalues of
$\left[ \partial^2V/\partial\chi_k\partial\chi_l\right] $
in fact requires that the $\{n_k\}$ be even:
\hbox{$n_k = 2 p_k$}, where
the $\{p_k\}$ are integers.

Hence we look for asymptotic solutions,
for
\ $\displaystyle {e_c^2\over 2 \pi \MM^2}\ll 1$,
of the form
$$ \sqrt{4\pi}\chi_k(\infty)= 2 p_k \pi \,+\, c_k {e_c^2 \over
2 \MM^2} \,+\, {\cal O}\left( {e_c^4\over \MM^4}\right)
\qquad\left( k=1,\dots,N_c\right)
\eqn\JI$$
The equations \SIX\ in fact tell us that for integer  $\{p_k\}$
the corrections $\{c_k\}$ are given by
$$c_k = 2 \left( -p_k \,+\, {1\over N_c}\left( \sum_{j=1}^{N_c}
p_j \right) \right) \qquad\left( k=1,\dots,N_c\right)
\eqn\JII$$
Equation \SVII\
then tells us that the baryon number
$$B={1\over\sqrt{\pi}} \sum_{k=1}^{N_c} \chi_k(\infty)
= \sum_{k=1}^{N_c} p_k \eqn\JIII$$
as $\displaystyle \sumk c_k = 0$. Hence $B$ is an integer.
The variable $\xi$ \SXI\ for the case
\hbox{$N_c{=}2$} may be generalized to the variables
$$\xi_k \equiv \sqrt{4\pi}\chi_k(\infty) - \left( \sum_{l=1}^{N_c}
p_l \right) \pi\qquad\left( k=1,\dots,N_c\right) \eqn\JIV$$
which take the values
$$\xi_k=(2p_k-B)\pi - {e_c^2\over \MM^2} \left( p_k - {1\over N_c}
\sum_{l=1}^{N_c} p_l\right)
+ \,\O\left({ e_c^2 \over \MM^2}\right)^2
\qquad\left( k=1,\dots,N_c\right)
\eqn\JV$$
generalizing the expression \SXII\ for $N_c{=}2$.
Correspondingly, the string tension \SXIII\ becomes
$$ \alphaprime
=2\pi e_c^2 \,\sum_{k=1}^{N_c} (B/N_c - p_k)^2\eqn\JVI$$
We see that when $B$ is some integer
 multiple of $N_c$, corresponding to
a multiple-baryon state,\refmark{\FZ}
the string tension vanishes,
$\alphaprime=0$, if all the $\{p_k\}$ are taken equal to the
multiple-baryon number.
\chapter{Finite temperature}
Further insight into the interpretation of these QCD$_2$ solitons can be
gained by considering the bosonized form of the action \SI,\SIV\ at
finite temperature. The temperature-dependent one-loop corrections to a
generic 2-dimensional action can be written in the form\refmark{\DJ}
$$\delta V(T) = {T\over 2 \pi} \int_0^\infty d k
\,\Tr\, \ln  \left( 1 - e^{-\beta \sqrt{k^2 + \MMM^2}} \right)
\eqn\TI$$
where $T={1/\beta }$ is the temperature and \MMM\ is the mass
matrix. The leading field-dependent term in \TI\ has the form
$$\delta V_1(T) = c\, T\,\Tr \sqrt{\MMM^2} \eqn\TII$$
where $c$ is a coefficient independent of the specific theory.
In the case $N_c=2$, the finite-temperature correction \TII\
to the effective potential \SVI\ takes the form
$$
\eqalign{
\delta V_1(T) & = c \,T \biggl( \sqrt{2} \MM \biggl(
\sqrt{\cos \sqrt{4\pi} \chi_1 } +
\sqrt{\cos \sqrt{4\pi} \chi_2 }\biggr)
\crr
&+
 {e^2_c \over 8 \sqrt{2}\pi } \biggl(
{1\over \MM \sqrt{\cos \sqrt{4\pi} \chi_1 }} +
{1\over \MM \sqrt{\cos \sqrt{4\pi} \chi_2 }} \biggr) \biggr)
\crr}
\eqn\TIII$$
We now discuss some likely implications of this correction
whilst recognizing\refmark{\DJ}
that a complete discussion of the phase structure
of $QCD_2$ would require evaluating the effective potential
to all orders in the loop expansion.

As discussed in Section~2,
 ``baryonic" soliton solutions to the zero-temperature
$N_c=2$ field equations (even $B$) exist for arbitrary values of
$\epsilon=e_c^2/2\pi \MM^2$, whilst ``quark" (odd $B$) solitons
exist only for $\epsilon < 1$.
As follows from equations \SXI,\SXII,
they correspond to
$$\chi_1(\infty)\simeq m \sqrt{\pi} + \epsilon \sqrt{\pi}
\left( {B\over2}-m\right) \qquad \left(m \in \ZZ\right)
\eqn\TIV$$
in the limit $\epsilon\ll1$, and
\hbox{$\chi_2(\infty) = -\chi_1(\infty) + n \sqrt{\pi}$}.
In the case of even $B$, there are baryonic solitons with
$m=B/2$ for which the $\epsilon$-dependent correction in
$\TIV$ vanishes, whereas it is always present for quark
solitons with odd $B$, and indeed becomes large when
\hbox{$\epsilon\rightarrow 1$},
 corresponding to the previously noted disappearance
of the soliton.
When $\epsilon\ll 1$, the leading effect of the finite-temperature
correction \TIII\ is to replace the zero-temperature solution
\TIV\ by
$$ \chi_1^T(\infty) \simeq m \sqrt{\pi}
+ \epsilon(T) \sqrt{4\pi}\left( {B\over2}-m\right)
\eqn\TV$$
where
$$\epsilon(T) \simeq \epsilon / \left( 1 - {c T \over
\MM \sqrt{2}}\right) \eqn\TVI$$
The effect of the finite-temperature
modification \TVI\ is in the direction of destabilizing the
quark solitons (and also baryonic solitons with $m\neq B/2$),
which presumably disappear at critical temperatures
${T_c}_{B,m} = {\cal O}(\MM)$, with solitons whose values of
$\vert B/2 - m \vert$ are largest disappearing first. In particular,
the last to disappear would be the quark solitons
with $\vert B/2 - m \vert = {1\over2}$, and finally the
baryons with $B=m$. The latter actually disappear only
when the coefficient of the periodic contribution
to the zero-temperature potential \SVI\ is finally
overwhelmed by the finite-temperature corrections
\TIII, which occurs at
${T_c}_{b,B/2}=T_{q/h}={\cal O}(\MM)$.

We now discuss briefly some possible interpretations
of these results. We interpret $T_{q/h}\simeq{\cal O}(\MM)$ as the
quark/hadron phase transition temperature, where the
fermion condensates dissolve,\refmark{\BG,\GLtemp},
the baryonic solitons disappear,
and correspondingly quarks
are deconfined, as discussed in refs. [\CEOI],[\CEOII].
We are not surprised by the fact that $T_{q/h}$ seems
to be quark mass- (and hence flavor-) dependent
in QCD$_2$, whereas it is expected to be ${\cal O}(\Lambda_{QCD})$
and flavor-independent in QCD$_4$. In QCD$_4$, $T_{q/h}$
can be determined by entropy considerations (the flux tube fluctuates
more and more until flux lines eventually fill all space), whereas
in QCD$_2$ the flux line cannot fluctuate out of the single space
dimension, and $T_{q/h}$ is determined by the energetics of quark
pair-creation, which becomes favoured when $T={\cal O}(\MM)$. The
presence of fermion condensation is \undertext{necessary} for
solitons to form, but not \undertext{sufficient}. Even when
$T=0$, for any fixed value of $\epsilon $
solitons with large values of
$\vert {B\over2}- m\vert$ do not exist and the number of
such solitons diminishes as $T$ increases, with the ``quark"
solitons disappearing before the $B=2m$ ``baryons".
\comm{Why do quarks disappear before baryons ?}

We plan to return to these and other issues at finite
temperature and chemical potential in a future
publication\refmark{\EFKb}

\chapter{Conclusions}

We have shown in this paper that QCD$_2$ has quark soliton solutions
if the quark mass is sufficiently large. We have also shown how these
quark solitons disappear when the quark mass $m_Q$
is reduced until the meson mass $\MM$ \SIIa\
becomes comparable to the dimensional gauge coupling strength $e_c$,
or when the temperature is increased to ${\cal O}(\MM)$ in the approach
to the quark/hadron phase transition. The next step in this approach
to the derivation of constituent quarks from QCD is to look for meson
solutions to the QCD$_2$ field equations which contain one heavy
and one light quark. The light meson cloud in the presence
of a heavy quark would correspond to the concept of a light
constituent quark in QCD.\refmark{\EFHK}
Once these solutions are obtained, we hope to be
able to abstract the relevant features of the field equations in
QCD$_4$ and then solve them to construct constituent quark
solitons also in 4 dimensions.
\comm{Why is \MM\ the quark mass here? We defined it as the meson
mass scale?}

\ACK
Y.F. wishes to thank the Aspen center for hospitality during
the period when this work was completed.
The research of M.K.  was supported in part
by the Basic Research Foundation administered by the
Israel Academy of Sciences and Humanities
and by grant No.~90-00342 from the United States-Israel
Binational Science Foundation(BSF), Jerusalem, Israel.

\refout

\par\penalty-400
   \vskip\chapterskip\spacecheck\referenceminspace
   \line{\fourteenrm\hfil FIGURE CAPTION\hfil}\vskip\headskip

\noindent
    Comparison of the left- and right-hand sides of the soliton
eq.~\SIXa,
applying the relation \SXa\
corresponding to a quark soliton with $n=1$, for
$\epsilon = e_c^2/2\pi \MM^2= 0.1$ (solid line) and
$\epsilon =0.5$ (dotted line). There are no solutions for
$\epsilon >1$
(dot-dashed line, drawn for $\epsilon =2$).

\bye